\documentclass{article}
\usepackage{graphicx}
\usepackage{amssymb}
\usepackage{xcolor}
\usepackage{subfigure}
\usepackage{amsmath}
\usepackage[letterpaper,left=2.5cm]{geometry}
\begin{document}
\title{New approaches in agent-based modeling of complex financial systems}


\maketitle
\author{ \centering
T. T. Chen$^{1,2}$, B. Zheng$^{1,2,*}$, Y. Li$^{1,2}$, X. F. Jiang$^{1,2}$ \\
$^{1}$Department of Physics, Zhejiang University, Hangzhou 310027, P.R. China\\
$^{2}$Collaborative Innovation Center of Advanced Microstructures, Nanjing 210093, P.R. China\\
$*$Corresponding authors.\ E-mail: $^*$zhengbo@zju.edu.cn \\}


\abstract{Agent-based modeling is a powerful simulation technique to understand the collective behavior and microscopic interaction in complex financial systems.
Recently, the concept for determining the key parameters of the agent-based models from empirical data instead of setting them artificially was suggested.
We first review several agent-based models and the new approaches to determine the key model parameters from historical market data.
Based on the agents' behaviors with heterogenous personal preferences and interactions, these models are successful to explain
the microscopic origination of the temporal and spatial correlations of the financial markets. We then present a novel paradigm combining the big-data analysis with the agent-based modeling.
Specifically, from internet query and stock market data, we extract the information driving forces, and develop an agent-based model to simulate
the dynamic behaviors of the complex financial systems.
}


\section{Introduction}
\noindent Complex financial systems typically have many-body interactions. The interactions of multiple agents induce various collective phenomena, such as the abnormal distributions, temporal correlations, and sector structures \cite{bla76,man95,gop99,liu99,gab03,qiu06,jia14,zhe14,tan15}. The complex financial systems are also substantially influenced by the external information that may, for example, drive the systems to non-stationary states, larger fluctuations or extreme events \cite{pre11,pod11,li11,tum12,mun12,jia12,jia13,yur14}.

Complex financial systems are important examples of open complex systems.
Standard finance supposes that investors have complete rationality, but the progress of behavioral finance and experimental fiance shows that investors in the real life have behavioral and emotional differences \cite{koo02,jo08}. To be more specific, agents who are not fully rational may have different personal preferences and interact with each other in financial markets \cite{fen12,che13,gon14,sha14,che15,kai15,sav15,tan15}.

Information is a leading factors in complex financial systems.
However, our understanding of external information and its controlling effect in the agent-based modeling is rather limited \cite{sam07,man11,cha11b,sor14}.
In recent years, exploring the scientific impact of online big-data has attracted much attention of researchers from different fields. The massive new data sources resulting from human interactions with the internet offer a better understanding for the profound influence of external information on complex financial systems \cite{pre12b,bor12,pre13,moa13,his13,kri13,nog14,cur14,lil15}.

Agent-based modeling is a powerful simulation technique to understand the collective behavior in complex financial systems \cite{gia01,bon02,eba04,ren06a,far09,sch09}. More recently, the concept for determining the key parameters of the agent-based models from empirical data instead of setting them artificially was suggested \cite{fen12}. Similar concept has also been applied to the order-driven models, which were first proposed by Mike and Farmer \cite{mik08} and improved by Gu and Zhou \cite{gu09a,gu09b,men12,zho16}. In this family of order-driven models, the parameters of order submissions and order cancellations are determined using real order book data. For comparison, the agent-based models focus more on the behaviors of agents \cite{gia01,bon02,eba04,ren06a,far09,sch09}, while the order-driven models are mainly intended to explore the dynamics of the order flows \cite{mik08,gu09a,gu09b,men12,zho16}. In section 2, we review several agent-based models that are based on the agents' behaviors with heterogenous personal preferences and interactions. These models explore the microscopic origination of the temporal and spatial correlations of the financial markets \cite{che13,che15,tan15}.
In section 3, we present a novel paradigm combining the big-data analysis with the agent-based modeling \cite{che16}.

\section{New approaches in agent-based models}

\noindent From the view of physicists, the dynamic behavior and community structure of complex financial systems can
be characterized by temporal and spatial correlation functions. Recently, several agent-based models are proposed to explore the microscopic generation mechanisms of temporal and spatial correlations \cite{che13,che15,tan15}. These models are microscopic herding models, in which the agents are linked with each other and trade in groups, and in particular, contain new approaches to multi-agent interactions.

\subsection{Basics of agent-based models}

\noindent The stock price on day $t$ is denoted as $Y(t)$, and the logarithmic price return is $R(t)=\ln [Y(t)/Y(t-1)]$. For comparison of different time series of returns, the normalized
return $r(t)$ is introduced,
\begin{equation}
r(t)=[R(t)-\langle R(t)\rangle]/\sigma,\label{eq:norm}
\end{equation}
where $\langle \cdots\rangle $ represents the average
over time $t$, and $\sigma=\sqrt{\langle R^{2}(t)\rangle -\langle R(t)\rangle ^{2}}$ is the standard deviation of
$R(t)$.
In stock markets,
the information for investors is highly incomplete, therefore
an agent's decision of \emph{buy}, \emph{sell} or \emph{hold} is assumed to be random. In these models, there is only one stock and there are $N$ agents, and each operates one share every day.
On day $t$, each agent $i$ makes a
trading decision $\phi_{i}(t)$,
\begin{eqnarray} \phi_{i}(t)=\left\{
\begin{array}{rcl}
1 ~~     &      & \textnormal{buy}\\
-1~~    &      & \textnormal{sell}\\
0 ~~    &      & \textnormal{hold}
\end{array} \right. 
\end{eqnarray}

and the probabilities of
\emph{buy}, \emph{sell} or \emph{hold} decisions are denoted as $P_{buy}(t)$, $P_{sell}(t)$ and $P_{hold}(t)$, respectively.
The price return $R(t)$ is defined by the difference of the demand and supply of the stock,
\begin{equation}
R(t)=\sum_{i=1}^{N}\phi_{i}(t).\label{eq:return}
\end{equation}
For simplicity, the volatility is defined as the absolute return $|R(t)|$. Other definitions yield similar results.

The investment horizon is introduced since agents' decision is based on the previous stock performance
of different time horizons \cite{che13,che15,tan15}. It has been found
that the relative portion $\gamma_{i}$ of agents with an $i$-days investment
horizon follows a power-law decay, $\gamma_{i} \propto i^{- \eta}$
with $\eta=1.12$ \cite{fen12}. The maximum investment horizon is denoted as $M$. To describe the integrated investment basis of all agents, a weighted average return $R'(t)$ is introduced,
\begin{equation}
R'(t)=k\cdot\sum_{i=1}^{M}\left[\gamma_{i}\sum_{j=0}^{i-1}R(t-j)\right],\label{eq:fR}
\end{equation}
where $k$ is a proportional coefficient.  According to Ref.~\cite{men10}, the investment horizons of investors range from a few days to several months.  For $M$ between $50$ and $500$, the results from simulations remain robust.

In complex financial systems, herding is one of the collective behaviors,
which arises when investors imitate the decision of others rather than follow their own belief and judgement. In other words, the investors cluster into groups when making decisions\cite{egu00,ken11}. Here a herding degree $D(t)$ is introduced to quantify the clustering degree of the herding behavior,
\begin{equation}
D(t)=n_A(t)/N,\label{eq:dt}
\end{equation}
where $n_A(t)$ is the average number of agents in each cluster on day $t$.

\subsection{Agent-based model with asymmetric trading and
herding}\label{ASModel}

\noindent The negative and positive return-volatility correlations, i.e., the so-called
leverage and anti-leverage effects, are particularly important for the understanding of the price dynamics \cite{bla76,qiu06,she09a,par11}. Although various
macroscopic models have been proposed to describe the return-volatility correlation, it is very important to understand the correlations from the microscopic level.
To study the microscopic origination of the return-volatility correlation in financial markets, two novel microscopic mechanisms, i.e., investors'
asymmetric trading and herding in bull and bear markets, are recently introduced in the agent-based modeling \cite{che13}.

\textbf{1. Two important behavior of investors}

\textbf{(i) Asymmetric trading.}
An investor's willingness to trade is
affected by the previous price returns, leading the trading probability to be distinct in bull and bear markets.
The model thus assumes dynamic probabilities for buying and selling, but with $P_{buy}(t)=P_{sell}(t)$.
As the trading probability $P_{trade}(t)=P_{buy}(t)+P_{sell}(t)$, its average over time is set to be $\langle P_{trade}(t)\rangle =2p$.
We adopt the value of $p$ estimated in Ref. \cite{fen12}, $p=0.0154$. The market performance is defined to be bullish if $R'(t)>0$,
and bearish if $R'(t)<0$.
The investors' asymmetric trading in bull and bear markets
gives rise to the distinction between $P_{trade}(t+1)|_{R'(t)>0}$ and $P_{trade}(t+1)|_{R'(t)<0}$.
Thus, $P_{trade}(t+1)$ should take the form

\begin{equation}
\begin{array}{rcl}
P_{trade}(t+1)=2p\cdot\alpha & \; R'(t)>0 \\
P_{trade}(t+1)=2p & \; R'(t)=0\\
P_{trade}(t+1)=2p\cdot\beta & \; R'(t)<0
\end{array}\label{eq:fp}.
\end{equation}
Here $\alpha$ and $\beta$ are asymmetric factors, and $\langle P_{trade}(t)\rangle =2p$ requires $\alpha+\beta=2$, i.e., $\alpha$ and $\beta$ are not independent.

\textbf{(ii) Asymmetric herding.}
Herding, as one of the collective behavior in financial markets, describes the fact that investors form cluster when making decisions, and these clusters can be large \cite{egu00,ken11}. Actually, the herding behavior in bull markets is not the same as that in bear ones \cite{kim05,wal06}.

In general, herding should be related to previous volatilities \cite{con00,bla12}, and we set the average number of agents in each cluster $n_A(t+1)=|R'(t)|$. Hence
the herding degree on day $t+1$ is
\begin{equation}
D(t+1)=|R'(t)|/N.\label{eq:od}
\end{equation}
This herding degree is symmetric for $R'(t)>0$ and $R'(t)<0$.
However, investors' herding behavior in bull and bear markets are asymmetric. Thus $D(t+1)$ should be redefined to
be
\begin{equation}
D(t+1)=|R'(t)-\Delta R|/N.\label{eq:fd}
\end{equation}
Here $\Delta R$ is the degree of asymmetry.
Everyday, the agents in a cluster make a same trading decision,
i.e., \emph{buy}, \emph{sell} or \emph{hold} with a same probability $P_{buy}$, $P_{sell}$ or $P_{hold}$.

\textbf{2. Determination of $\alpha$ and $\Delta R$}

Six representative stock-market indices are collected, i.e., the daily data for the S\&P 500 Index, Shanghai
Index, Nikkei 225 Index, FTSE 100 Index,
HKSE Index, and DAX Index.

We assume that the trading probability is proportional to the trading volume.
Thus the ratio of the average trading volumes for the bull markets and the bear ones is
\begin{equation}
V_{+}/V_{-}=\frac{P_{trade}(t+1)|_{R'(t)>0}}{P_{trade}(t+1)|_{R'(t)<0}}=\alpha/\beta.
\end{equation}
Together with the condition $\alpha+\beta=2$, $\alpha$ is determined from $V_{+}/V_{-}$ for the six representative stock market indices, as shown in Table~\ref{tab1}.

From empirical analysis, the herding degrees of bull and bear stock-markets are not equal, i.e., $d_{bull}\neq d_{bear}$.
To quantize this asymmetry, a shifting $\Delta r$ is introduced such that $d_{bull}[r'(t)]=d_{bear}[r'(t)]$ with $r'(t)=r(t)+\Delta r$. From this definition, $\Delta r$ is derived to be
\begin{equation}
\Delta r=\frac{1}{2}[d_{bear}(r(t))-d_{bull}(r(t))].
\end{equation}
Here the herding degrees of bull markets ($r(t)>0$) and bear markets ($r(t)<0$) are defined as the average $|r(t)|$ with the weight $V(t)$, i.e.,
\begin{equation}
\left\{ \begin{array}{c}
d_{bull}[r(t)]=\sum_{t,r(t)>0}V(t)\cdot r(t)/\sum_{t,r(t)>0}V(t)\\
d_{bear}[r(t)]=\sum_{t,r(t)<0}V(t)\cdot|r(t)|/\sum_{t,r(t)<0}V(t)
\end{array}\right.\label{eq:reald}.
\end{equation}
Then the shifting to the time series $R(t)$, which equalize the herding degree $D(t+1)=|R'(t)-\Delta R|/N$ in bull markets ($R'(t)>0$) and bear markets ($R'(t)<0$) is similarly computed. Table~\ref{tab1} shows the values of $\Delta r$ and $\Delta R$ for different indices.

\begin{table}[b]\small
\caption{\textbf{The values of  $\alpha$, $\Delta r$ and $\Delta R$
for the six indices.} $\Delta R$ is computed from the linear relation between $\Delta r$ and $\Delta R$ for all these indices.}
 \centering
\begin{tabular}{|c|c|cc|}
\hline
Index & $\alpha$ & $\Delta r$ & $\Delta R$\tabularnewline
\hline
S\&P 500 &  $1.01\pm0.01$ & $0.067\pm0.007$  & $3$\tabularnewline
Shanghai &  $1.09\pm0.01$ & $-0.043\pm0.005$ & $-2$\tabularnewline
Nikkei 225  & $1.01\pm0.01$ & $0.039\pm0.005$ & $2$\tabularnewline
FTSE 100 & $0.99\pm0.01$ & $0.028\pm0.003$ & $2$\tabularnewline
Hangseng  &  $1.02\pm0.02$ & $0.032\pm0.003$ & $2$\tabularnewline
DAX  &  $0.98\pm0.02$ & $0.013\pm0.002$ & $1$\tabularnewline
\hline
\end{tabular}
\label{tab1}
\end{table}

\textbf{3. Simulation results}

\begin{figure}[h]
\begin{center}
\includegraphics[width=0.45\textwidth]{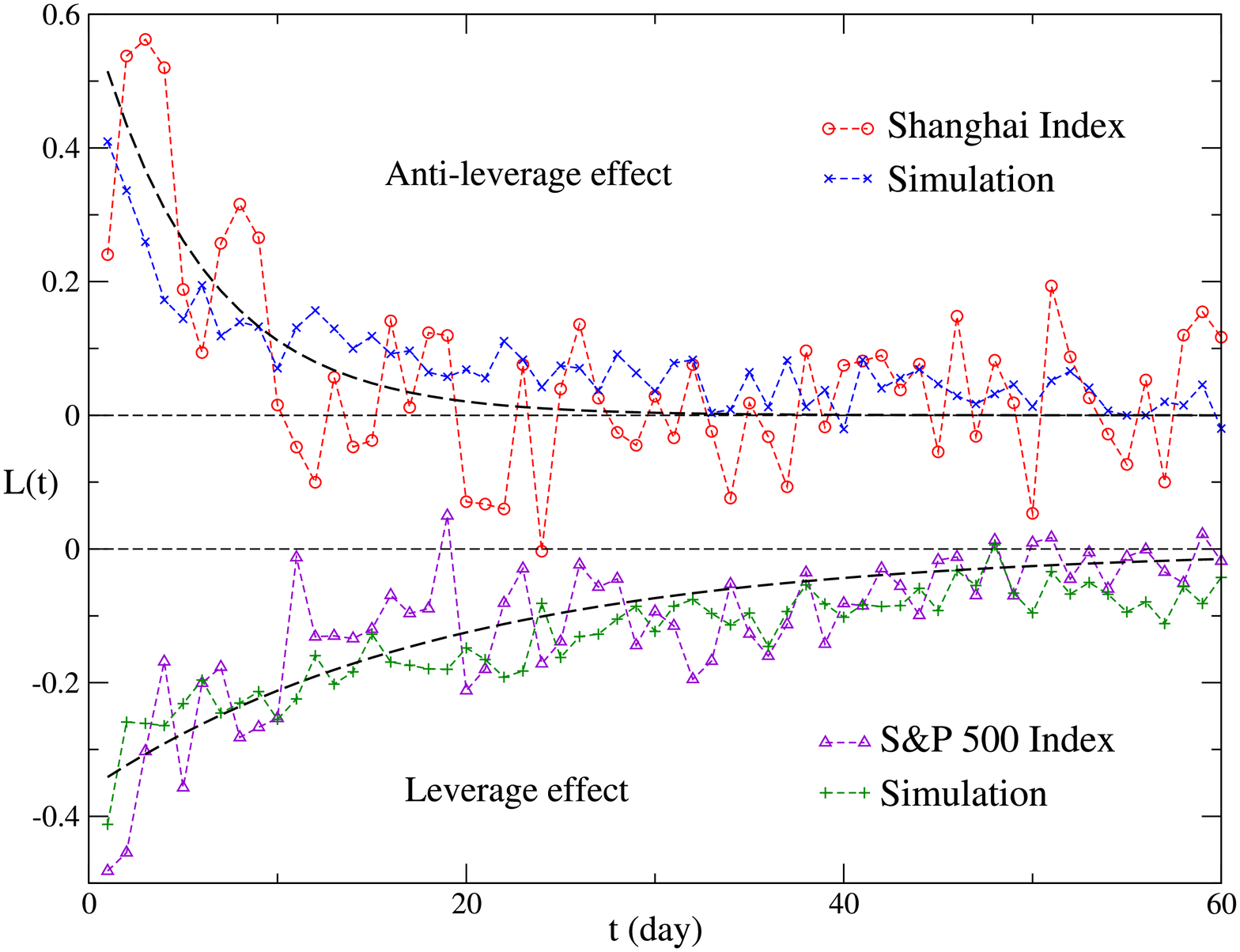}
\end{center}
\protect \caption{The return-volatility correlation functions for the S\&P 500 and Shanghai indices, and for the corresponding simulations. The S\&P 500 and Shanghai indices are simulated
with $(\alpha,\Delta R)=(1.0,3)$ and $(\alpha,\Delta R)=(1.1,-2)$, respectively. Dashed lines show the exponential fits $L(t)=c\cdot exp(-t/\tau)$.
}
\label{fig1}
\end{figure}

\begin{figure}[h]
\begin{center}
\includegraphics[width=0.65\textwidth]{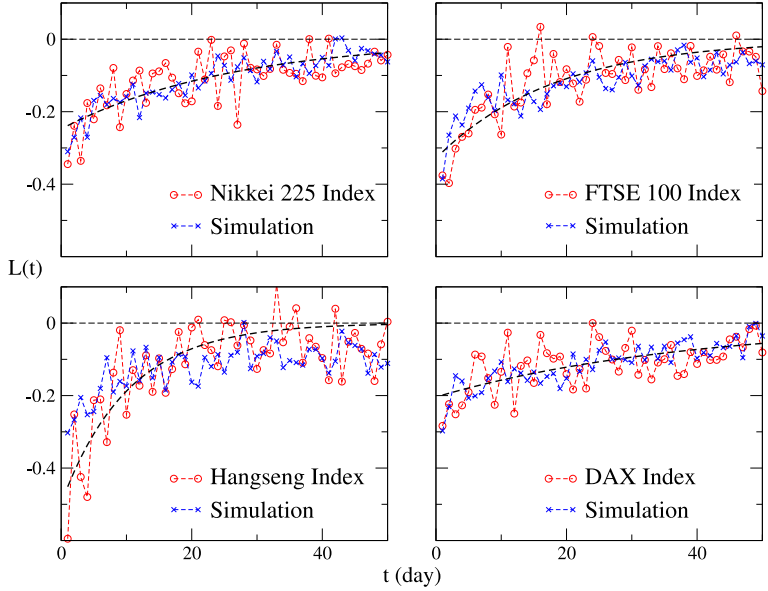}
\end{center}
\protect\caption{The return-volatility correlation functions for the four indices and the corresponding simulations.The Nikkei 225, FTSE 100, Hangseng and DAX indices are simulated
with $(\alpha,\Delta R)=(1.0,2)$, $(1.0,2)$, $(1.0,2)$ and $(1.0,1)$, respectively. Dashed lines show the exponential fits $L(t)=c\cdot exp(-t/\tau)$.}
\label{fig2}
\end{figure}

With $\alpha$ and $\Delta R$ determined for each index, the model produces the time
series of returns $R(t)$.
To describe how past returns affect future volatilities, the return-volatility correlation function $L(t)$ is defined,
\begin{equation}
L(t)=\langle r(t')\cdot|r(t'+t)|^{2}\rangle  /Z,
\label{L}
\end{equation}
with $Z=\langle |r(t')|^{2}\rangle ^{2}$ \cite{bou01}. Here $\langle \cdots\rangle $ represents the average
over time $t'$.
As displayed in Fig.~\ref{fig1}, $L(t)$ calculated with the empirical data of the S\&P 500 Index shows negative values up to at least 15 days, and this is the well-known leverage effect \cite{bou01,bla76,qiu06}. On the other hand, $L(t)$ for the Shanghai Index remains positive for about 10 days. That is the so-called anti-leverage effect \cite{qiu06,she09a}. The return-volatility correlation function  produced in the model is in agreement with that calculated from empirical data on amplitude and duration for both the S\&P 500 and Shanghai indices. This is the first result that the leverage and anti-leverage effects are simulated with a microscopic model.
As displayed in Fig.~\ref{fig2}, $L(t)$ for the simulations is also in agreement with that for the the Nikkei, FTSE 100, HKSE and DAX indices.

As shown in Fig. 4 and Fig. 5 of Ref. \cite{che13}, the model also produces the volatility clustering and the fat-tail distribution of returns \cite{che13}. The hurst exponent of $A(t)$ is calculated to be $0.79$, which also indicates the long-range correlation of volatilities \cite{sha12}. The auto-correlation function of returns fluctuates around zero. The power-law exponent of the simulated returns is estimated to be $2.96$, close to the so-called inverse cubic law \cite{gop98,gu08,ple99,mu10b}.

\subsection{Agent-based model with asymmetric trading preference}

\noindent The problem whether and how volatilities affect the price movement draws much attention. However, the usual volatility-return correlation function, which is local in time, typically fluctuates around zero. Recently, a dynamic observable nonlocal in time was constructed  to explore the volatility-return correlation \cite{tan15}. Strikingly, the correlation is found to be non-zero, with an amplitude of a few percent and a duration of over two weeks. This result provides compelling evidence that past volatilities nonlocal in time affect future returns. Alternatively, this phenomenon could be also understood as the non-stationary dynamic effect
of the complex financial systems.

To study the microscopic origin of the nonlocal volatility-return correlation, an agent-based model is constructed \cite{tan15},
in which a novel mechanism, i.e., the asymmetric trading preference in volatile and stable markets, is introduced.

In financial markets, the market behavior of buying and selling are not always in balance \cite{ple03}. Hence, $P_{buy}$ and $P_{sell}$ are not always equal to each other. They are affected by previous volatilities,
and the more volatile the market is, the more $P_{buy}$ differs from $P_{sell}$.

For an agent with an $i$-days investment horizon, the average volatility over previous $i$ days is taken into account, which is defined as
\begin{equation}
v_{i}(t)=\frac{1}{i}\sum_{j=1}^{i}v(t-j+1).
\end{equation}
The background volatility is considered to be $v_{M}(t)$ with $M$ being the maximum investment horizon. On day $t$, the agent with an $i$-days
investment horizon estimates the volatility of the market by comparing $v_{i}(t)$ with $v_{M}(t)$. Therefore, the integrated perspective of all agents
on the recent market volatility is defined as
\begin{equation}
\xi(t)=\frac{1}{v_{M}(t)}\sum_{i=1}^{M}\gamma_{i}v_{i}(t).
\end{equation}
Thus, the probabilities of buying and selling are assumed to be
\begin{equation}
\left\{ \begin{array}{l}
P_{buy}(t+1)=p[c\cdot\xi(t)+(1-c)]\\
P_{sell}(t+1)=2p-P_{buy}(t+1)
\end{array}\right..
\label{eq:fp}
\end{equation}
Here the parameter $c$ measures the degree of agents' asymmetric trading preference in volatile and stable markets. Compared with the model reviewed in Sec.~\ref{ASModel}, $c$ is the only additional parameter.
In principle, $c$ could be determined from the trade and quote data of stock markets. Unfortunately, the data are currently not available to us.
Thus the question how to determine $c$ from the historical market data remains open.
Anyway, with this model, it is possible to simulate the non-zero volatility-return correlation nonlocal in time \cite{tan15}.

\subsection{Agent-based model with multi-level herding}

\noindent The spatial structure of the stock markets is explored through the cross-correlations of individual stocks.
With the random matrix theory (RMT), for examples, communities can be identified, which are usually associated with business sectors in stock markets \cite{ple02,uts04,pan07,she09,pod10,jia12,jia14}.
To simulate the sector structure with the agent-based model, we newly introduce the multi-level herding mechanism \cite{che15}.

\begin{figure}[h]
\begin{center}
\includegraphics[width=0.7\textwidth]{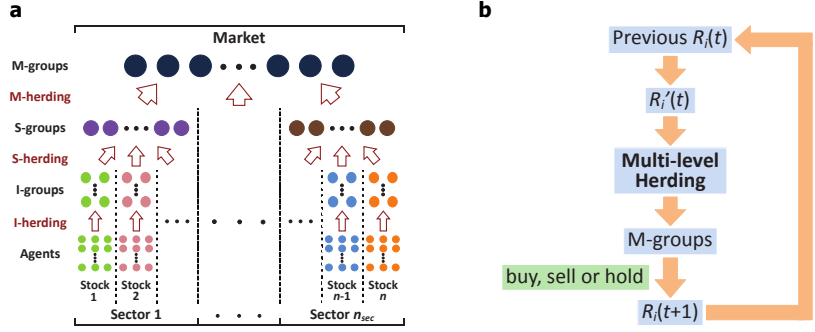}
\end{center}
\caption{The schematic diagram of (a) the multi-level herding; (b) the procedure of simulation.
}
\label{fig3}
\end{figure}

\textbf{1. Multi-level herding.} In the model, there are
$N$ agents, $n$ stocks and $n_{sec}$ sectors. Each sector contains $n/n_{sec}$ stocks. Every agent holds only one stock, which is randomly chosen from the $n$ stocks.
The logarithmic price return of the $k$-th stock on day $t$ is denoted by $R_{k}(t)$.
We assume that the agents' herding behavior comprises the herding at stock, sector and market levels. The schematic diagram of the multi-level herding is displayed in Fig.~\ref{fig3}(a).

(i) Herding at stock level. The agents in
each individual stock first cluster into groups, which are called I-groups. The herding degree $D^{I}$ quantifies the herding behavior at this level. On day $t$, the herding degree
for the $k$-th stock is
\begin{equation}
D_{k}^{I}(t)=|R_{k}'(t-1)|/N_{k}.\label{eq:HaI}
\end{equation}
In the $k$-th stock, the number of I-groups
is $1/D_{k}^{I}(t)$, and the agents randomly join in one of the
I-groups. After the herding at stock level,
the number of I-groups in the $j$-th sector and in the whole market
are, respectively, denoted by $N_{j}^{I}(t)$ and $N_{M}^{I}(t)$,
\begin{equation}
\left\{ \begin{array}{c}
N_{j}^{I}(t)=\sum\limits_{k\in j}[1/D_{k}^{I}(t)]\\
N_{M}^{I}(t)=\sum\limits_{k}[1/D_{k}^{I}(t)]
\end{array}.\right.
\end{equation}
Here $k\in j$ represents the stock $k$ in sector $j$.

(ii) Herding at sector level. The stocks in a same sector share the characteristics of the sector. At this level, agents' herding behavior is driven by the price co-movement of the sector, i.e., the prices of stocks in a sector tend to rise and fall simultaneously. Thus the I-groups in a same sector would further form
larger groups, which are called S-groups. $H_{M}$ and $H_{j}$ characterize the price co-movement degrees for stocks in the whole market and in sector $j$, respectively.
For the $j$-th sector, the average number of I-groups in each S-group is set
to be $n\cdot(H_{j}-H_{M})$, which represents the pure price co-movement of the
sector. Therefore the herding degree is
\begin{equation}
D_{j}^{S}(t)=n\cdot(H_{j}-H_{M})/N_{j}^{I}(t).\label{eq:HaS}
\end{equation}
In sector $j$, the number of S-groups is $1/D_{j}^{S}(t)$, and each I-group joins in one of the S-groups.

(iii) Herding at market level. Agents' herding behavior at this level is driven by the price co-movement of the entire market. The S-groups in different sectors share common features of the whole market, and thus cluster into larger groups. These groups are called M-groups. For the S-groups in sector $j$, the herding degree at market level is
\begin{equation}
D_{j}^{M}(t)=n \cdot H_{M}/N_{j}^{M}(t),
\end{equation}
and the number of M-groups is $1/D_{j}^{M}(t)$. The total number of M-groups in the market is the maximum of $1/D_{j}^{M}(t)$ for different $j$. With all M-groups numbered, an S-group in sector $j$ joins in one of the first $1/D_{j}^{M}(t)$ M-groups.

In the formation of S-groups, the I-groups in a
same stock tend not to join in a same S-group, otherwise these I-groups would have gathered together during the herding
at stock level. Similarly, in the formation of M-groups, the S-groups in a
same sector tend not to join in a same M-group.

After the herding for the three levels, all agents cluster
into M-groups. The agents in a same M-group make
a same trading decision $\phi_i(t)$ with a same probability.
The same as the previous models \cite{fen12,che13}, the buying and selling probabilities are equal, i.e., $P_{buy}=P_{sell}=P$, thus $P_{hold}=1-2P$.
Here $P$ is the buying or selling probability of an M-group, which can be calculated from the daily trading probability $p$ for each agent and the average number of agents in an M-group \cite{che15}.
The return of the $k$-th stock is defined as
$R_{k}(t)=\sum_{i \in k}\phi_{i}(t).\label{eq:R}$
Here $i \in k$ represents the agent $i$ in stock $k$.

\textbf{2. Determination of $H_{M}$ and $H_{j}$.}
On each day $t$, according to the sign of $r_{k}(t)$, the
stocks are grouped into two market trends, i.e., the rising and falling.
The amplitudes of the rising and falling trends on day $t$ are defined as
$v^{+}(t)$ and $v^{-}(t)$, respectively,
\begin{equation}
\left\{ \begin{array}{c}
v^{+}(t)=\sum_{i,r_{i}(t)>0}r_{i}^{2}(t)/n_{s}\\
v^{-}(t)=\sum_{i,r_{i}(t)<0}r_{i}^{2}(t)/n_{s}
\end{array}\right.
\end{equation}
Here $n_{s}$ is the number of stocks in a sector, and $n_{s}=n$ in the calculation of $H_{M}$. The amplitude $v^{d}(t)$ of the dominating trend and the amplitude $v^{n}(t)$ of the non-dominating one
are
\begin{equation}
\left\{ \begin{array}{c}
v^{d}(t)=max\{v^{+}(t),\: v^{-}(t)\}\\
v^{n}(t)=min\{v^{+}(t),\: v^{-}(t)\}
\end{array}\right.
\end{equation}
The stocks grouped into the dominating trend are denoted as the ``dominating
stocks''.

To characterize the price co-movement degrees for stocks in the whole market and in sector $j$, the co-movement degree $H_{M}$ and $H_{j}$ are computed,
\begin{equation}
\left\{ \begin{array}{c}
H_{M}=\left.\left\langle \zeta(t)\right\rangle \cdot\left\langle v^{d}(t)-v^{n}(t)\right\rangle \right|_{\textnormal{market}}\\
H_{j}=\left.\left\langle \zeta(t)\right\rangle \cdot\left\langle v^{d}(t)-v^{n}(t)\right\rangle \right|_{\textnormal{sector $j$}}
\end{array}.\right.
\end{equation}
where $|_{\textnormal{market}}$ and $|_{\textnormal{sector $j$}}$ represent the stocks in the whole market and in the $j$-th sector, respectively. $\zeta(t)$ represents the similarity in the signs of the returns for different stocks, and it is defined as the percentage of the dominating stocks, i.e., $\zeta(t)=n^{d}(t)/n_{s}$. $\left\langle v^{d}(t)-v^{n}(t) \right\rangle$ is the average total amplitude of the ``dominating stocks''.

The co-movement degree $H_{M}$ and $H_{j}$ for the NYSE and HKSE are shown in Table~\ref{tab2}.

\begin{table}[h]\small
\protect\caption{The values of parameters $H_{M}$ and $H_{j}$ for the NYSE and HKSE.
}
 \centering
\begin{tabular}{c|c|ccccc}
\hline
 & $H_{M}$ & $H_{1}$ & $H_{2}$ & $H_{3}$ & $H_{4}$ & $H_{5}$\tabularnewline
\hline
NYSE & 0.363 & 0.491 & 0.414 & 0.438 & 0.431 & 0.546\tabularnewline
HKSE & 0.306 & 0.426 & 0.406 & 0.364 & 0.361 & 0.340\tabularnewline
\hline
\end{tabular}
\label{tab2}
\end{table}

\begin{figure}[h]
\begin{center}
\includegraphics[width=0.5 \textwidth]{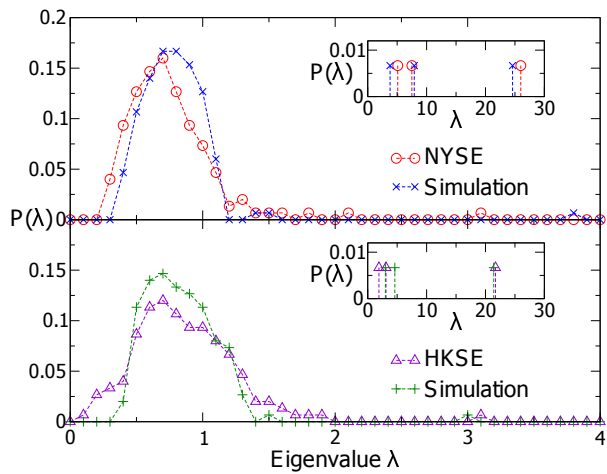}
\end{center}
\protect \caption{The probability distribution of the eigenvalues of the cross-correlation matrix $C$ for the NYSE and HKSE, and for the corresponding simulations. The inset shows the three largest eigenvalue for the NYSE and HKSE, and for the corresponding simulations.}
\label{fig4}
\end{figure}

\textbf{3. Simulation results.} Estimated from the historical market data and investment reports, the buying or selling probability is $P=0.363$ for the NYSE and $P=0.317$ for the HKSE \cite{che15}. With $H_{M}$ and $H_{j}$ determined for the NYSE and HKSE respectively,
the model produces the time series $R_{k}(t)$ of each stock. The schematic diagram of the simulation procedure is displayed in Fig.~\ref{fig4}(b).

To characterize the spatial structure, one may compute the equal-time cross-correlation matrix $C_{ij}=\langle r_{i}(t) r_{j}(t) \rangle$ \cite{she09,ple99},
where $\langle \cdots\rangle $ represents the average over time $t$, and $C_{ij}$ measures the correlation between the returns of the $i$-th and $j$-th stocks.
The distribution of the eigenvalues is displayed for the NYSE and HKSE in Fig.~\ref{fig4}, and the bulk of the distribution and the three largest eigenvalues from the simulation are in agreement with those from the empirical data.

\begin{figure}[h]
\begin{center}
\includegraphics[width=0.7\textwidth]{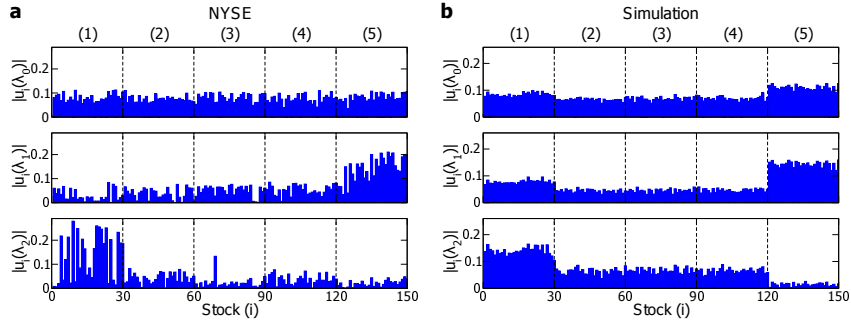}
\end{center}
\caption{The absolute values of the eigenvector components $u_{i}(\lambda)$ corresponding to the three largest eigenvalues for the cross-correlation matrix $C$ calculated from (a) the empirical data in the NYSE; (b) the simulated returns for the NYSE. Stocks are arranged according to business sectors separated by dashed lines. (1): Basic Materials; (2): Consumer Goods; (3): Industrial Goods; (4): Services; (5): Utility.
}
\label{fig5}
\end{figure}

The first, second and third largest eigenvalues of $C$ are denoted by $\lambda_{0}$, $\lambda_{1}$ and $\lambda_{2}$, respectively. $\lambda_{0}$ represents
the market mode, i.e., the price co-movement of the entire market, and the components of the corresponding
eigenvector is rather uniform for all stocks. Other large eigenvalues stand for the sector modes, and the eigenvector of these eigenvalues is dominated by the stocks in a certain sector.

The empirical result of the NYSE is displayed in Fig.~\ref{fig5}(a). The eigenvectors of $\lambda_{1}$ and $\lambda_{2}$
are dominated by sector $(5)$ and sector $(1)$ respectively, with the components significantly larger than those in other sectors.
These features are reproduced in our simulation, and the results are shown in Fig.~\ref{fig5}(b). For the HKSE, the eigenvectors
of $\lambda_{1}$ and $\lambda_{2}$ are respectively dominated by sector $(1)$ and sector $(2)$, and these features are also obtained \cite{che15}. From the simulated returns, we also observe the volatility clustering.

\section{Big data and agent-based modeling}\label{Model}
\noindent Information is one of the leading factors in complex financial systems.
In the past years, however, it is difficult to quantify the effect of the external information on the financial systems, due to the lack of data.
Our understanding of the external information and its controlling effect in the agent-based modeling is rather limited \cite{sam07,man11,cha11b,sor14}.
Fortunately, massive new data sources are resulted from human interactions with the internet in recent years.
Therefore, we propose a novel paradigm combining the big-data analysis with the agent-based modeling \cite{che16}.

\subsection{Information driving forces}\label{Force}
The internet query data can not only reflect the arrival of news, but also provide a proxy measurement of the information gathering process of the traders before their trading decisions.
We collect the weekly Google search volumes, and the corresponding historical market data for $108$ components of the S\&P 500. In this section, we define an information driving force and analyze how it drives the complex financial system.

The states of the external information, i.e., the Google search volume $G_k(t)$ for the $k$-th stock, may be complicated \cite{his13,lil15}.
As a first approach, we simplify the information states to two states, i.e.,
\begin{eqnarray}
S_k(t)=\left\{
\begin{array}{rcl}
1 ~~&& G_k(t)>\bar{G}_k \\
0 ~~&& G_k(t)\leq \bar{G}_k
\end{array} \right.
\end{eqnarray}
Here $\bar{G}_k$ is the mean value of $G_k(t)$.
The traders are more influenced by the external information at $S_k(t)=1$, while less at $S_k(t)=0$.
\begin{figure}[t!]
 \centering
\includegraphics[width=0.8\textwidth,clip]{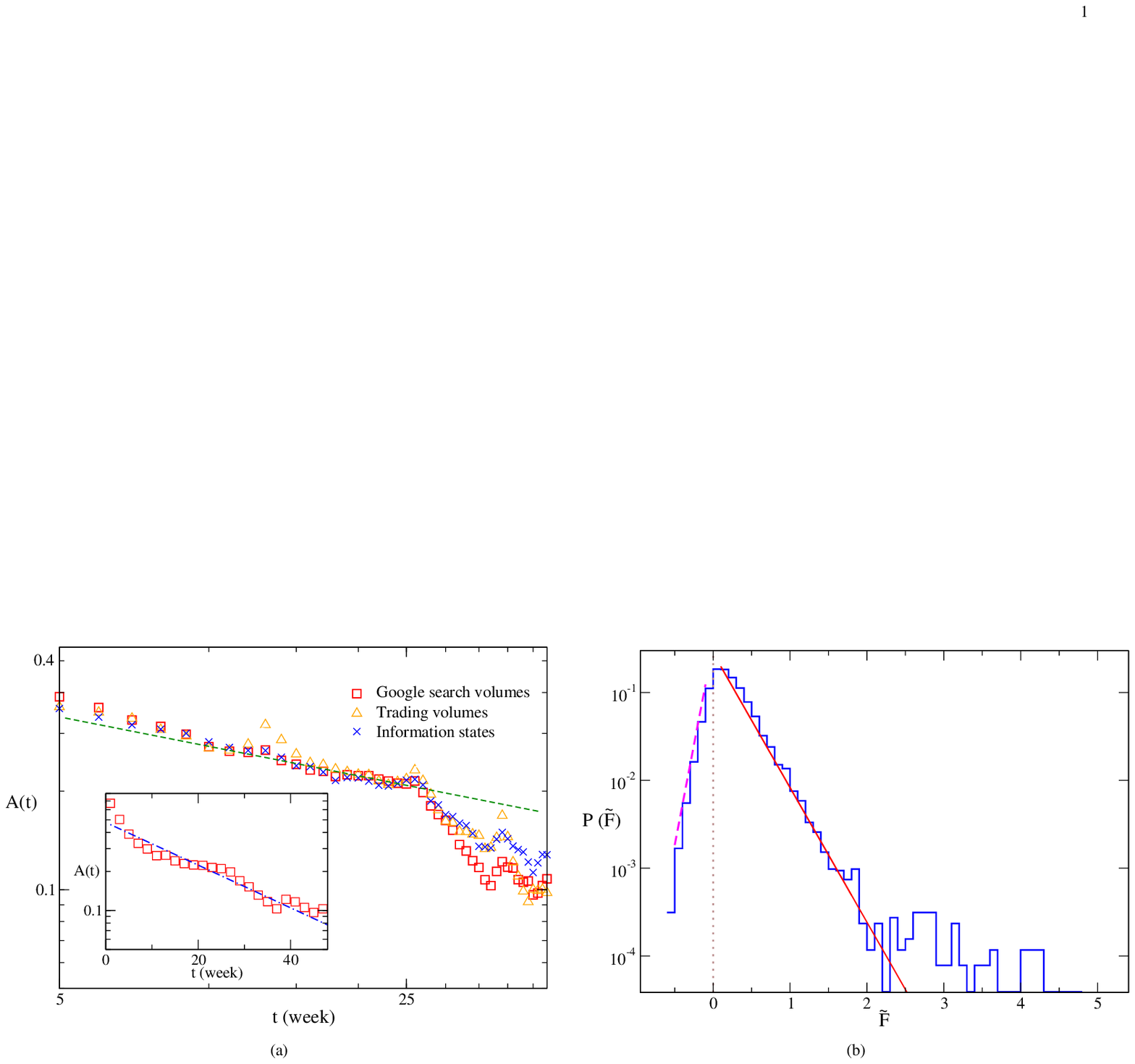}  
\protect\caption{(a) The average auto-correlation functions of the Google search volumes, trading volumes and the information states of the S\&P 500 components. 
A power-law fit is given by the dashed line. As shown in the inset, the curve is fitted with an exponential law $A(t) = c \exp(-t/\tau)$ with $\tau=26$.
(b) The probability distribution of the information driving forces for the S\&P 500 components. An exponential fit $P(\tilde{F}_k) = a_1 \exp(-b_1 \tilde{F}_k)$  with $b_1=3.5$ for $\tilde{F}_k(t)>0$ is displayed with the solid line.
The dashed line corresponds to an exponential fit for $\tilde{F}_k(t)<0$, i.e., $P(\tilde{F}_k) = a_2 \exp(b_2 \tilde{F}_k)$ with $b_2=10.5$.
\label{fig7}
}\end{figure}
The auto-correlation function of a time series $r(t')$ is defined as
\begin{eqnarray}
A(t)=[\langle|r(t')||r(t+t')|\rangle-\langle|r(t'|\rangle^{2}]/A_{0},
\label{At}\end{eqnarray}
where $A_{0}={\langle}|r(t')|^{2}\rangle-\langle|r(t')|\rangle^{2}$ \cite{che13}.
For each stock, the auto-correlation functions of $G_k(t')$, $S_k(t')$ and $V_k(t')$ are computed and averaged over $k$. As displayed in Fig.~\ref{fig7}(a), the average auto-correlation functions of $G_k(t')$ and $S_k(t')$
exhibit a power-law-like behavior in a certain period of time. This is similar to that of $V_k(t')$.
On the other hand, all the three curves start deviating from the power law at about $t=26$ weeks,
which could be considered as the correlating time $\tau$.

To study how the external information influences the trading behavior of the traders, we calculate the moving time averages of the trading volumes in different information states. Here we adopt the correlating time $\tau$ of the Google search volumes as the length of the moving time window. Denoting the moving time averages of the trading volumes at $S_k(t')=1$ and $S_k(t')=0$ by $V_k^1(t)$ and $V_k^0(t)$ respectively,
one can simply compute
\begin{eqnarray}
\begin{array}{rcl}
V_k^1(t)=\langle V_k(t') \rangle_{\tau}\mid_{S_k(t)=1},\\
V_k^0(t)=\langle V_k(t') \rangle_{\tau}\mid_{S_k(t)=0},
\end{array}
\end{eqnarray}
where $\langle\cdot\cdot\cdot\rangle_{\tau}$ represents the average over the time window $t' \in (t,t+\tau)$.
We then empirically define the information driving force for the $k$-th stock on time $t$
\begin{eqnarray}
  \tilde{F}_k(t)=V_k^1(t)/V_k^0(t)-1.
\end{eqnarray}
If $\tilde{F}_k(t)>0$, i.e., $V_k^1(t)>V_k^0(t)$, the traders trade more
frequently at the state $S_k(t)=1$, and the external information does drive the market to be more active. The positive information driving forces reflect the information gathering process of the traders before their trading decisions.
If $\tilde{F}_k(t)<0$, i.e., $V_k^1(t)<V_k^0(t)$, the traders trade less
frequently at the state $S_k(t)=1$, and the market is not driven to be more active.
The negative information driving forces may be related to the ambiguous or uncertain information which does not play a key role in the trading behavior.
As displayed in Fig.~\ref{fig7}(b), the probability distribution of $\tilde{F}_k(t)$ is obviously asymmetric with a heavier positive tail.
This result indicates that the external information usually drives the market to be more active,
which is consistent with the previous empirical findings for the internet query data or news \cite{his13,bor12}.

To study the information driving forces in different market states,
we compute the average information driving forces $\tilde{F}^{bear}$ and $\tilde{F}^{bull}$ for the bull and bear markets respectively. Thus their difference is defined as
\begin{eqnarray}
\Delta \tilde{F}= (\tilde{F}^{bear}-\tilde{F}^{bull})/{\langle \tilde{F}\rangle},
\label{dtheta}
\end{eqnarray}
where ${\langle \tilde{F} \rangle}$ is the mean value of $\tilde{F}_k(t)$ for all different $t$ and $i$.
The result is $\Delta \tilde{F}=0.4$, i.e., the information driving forces in the bear market are stronger than that in the bull market. The asymmetric information
driving forces in the bull market and bear market indicate that traders are more sensitive in the bear market.

\subsection{Agent-based model driven by information driving forces}
\textbf{1. Model framework.}\label{model framework}
As an application, we propose an agent-based model driven by the information driving force.
We consider a stock market composed of $N$ agents, in which there is only one stock, and each agent operates one share every day. On day $t$, after all the agents have made their trading decision $\phi_i(t)$ according to Eq.~(\ref{eq:si}), we can calculate the price return according to Eq.~(\ref{eq:fR}).
We still assume $P_i^{buy}(t)=P_i^{sell}(t)$, but the trading probability of the $i$-th agent $P_i(t)=P_i^{buy}(t)+P_i^{sell}(t)$ evolves with time.

The information driving force of the $i$-th agent in this section is denoted by $F_i(t)$, which is distinguished from the empirically-defined information driving force $\tilde{F}_k(t)$ of the $k$-th stock in section \ref {Force}.
We assume that $F_i(t)$ induces a dynamic fluctuation of the trading probability,
\begin{eqnarray}\label{main}
P_i(t)=E(1+F_i(t))P(0),
\end{eqnarray}
where $P(0)$ is the initial value of $P(t)$, and $E$ is the identity matrix.
In our model, we set $P(0)=2p/(1+\bar{F})$ to ensure the time average of the trading probabilities for $i$-th agent $\langle P_i(t) \rangle =2p$,
where $p=0.0154$ \cite{fen12}. Here $\bar{F}$ is the mean value of information driving forces $F_i(t)$.

\textbf{2. Information states.}\label{Statehistory}
As stated in section \ref{Force}, there are two information states for the market, i.e., $S(t)=1$ and $S(t)=0$, and the information driving force plays an important role only at the state $S(t)=1$. Here we omit the subscript $k$ of $S_k(t)$, for the difference between different stocks is not discussed in our model. The initial information state is randomly set to be $S(t)=1$ or $S(t)=0$. Then the information state will flip between $S(t)=1$ and $S(t)=0$ with an average transition probability $p_{t}$. On average, an information state will persist for $1/p_{t}$. Then we assume $p_{t}=1/\tau$, where $\tau$ is the correlating time of the Google search volume for the S\&P 500 components.

For simplicity, we only consider the positive information driving forces,
since the negative ones are not dominating. In section \ref{Force}, the probability distribution of the empirically-defined information driving forces are fitted with the exponential function
$Prob(\tilde{F}) \sim \exp(-b_1 \tilde{F})$ with $b_1=3.5$.
We suppose that $F_i(t)$ of different $i$ obeys
the same distribution. Therefore, the simplest form of $F_i(t)$ should be
\begin{eqnarray}\label{simpleft}
F_i(t)=s_i(t) y(t),
\end{eqnarray}
where the stochastic variable $y(t)$ obeys the distribution $Prob(y)$, and $s_i(t)$ is the state of the $i$-th agent. Each time $t$, we set the states for a dominating percentage of the agents to be $s_i(t)=S(t)$, and the states for the others to be $s_i(t)=1-S(t)$.

To describe the asymmetric trading behavior of agents in the bull and bear markets, we complete the form of $F_i(t)$,
\begin{eqnarray}\label{intensity}
F_i(t)=s_i(t) y(t)[1+a \cdot  sgn(R'(t))],
\end{eqnarray}
where $a$ is the asymmetric coefficient, and $R'(t)$ is the weighted return defined in Eq.~(\ref{eq:fR}).
We assume that the asymmetric coefficient $a=\Delta \tilde{F} /2$.
Here $\Delta\tilde{F}$ is the difference of the empirically-defined information driving forces in the
bull and bear markets, which is computed from Eq.~(\ref{dtheta}) in section \ref{Force}.

The herding behavior can be explained by the information dispersion \cite{egu00,cor07}. The agents behave similarly because they are exposed to the same information. Here we assume that the agents with positive information driving forces $F_i(t)$ are divided into clusters. The average number of agents in each cluster $n(t)$ should be related to the information driving force, and we set $n(t)=p_t^{-1}{\Sigma}_{i=1}^N F_i(t)/N$.

\textbf{3. Simulation results.}\label{Simulation}
With the number of agents set to be $N=10^4$, we perform the numerical simulation and obtain the return time series $R(t)$.

\begin{figure}[h]
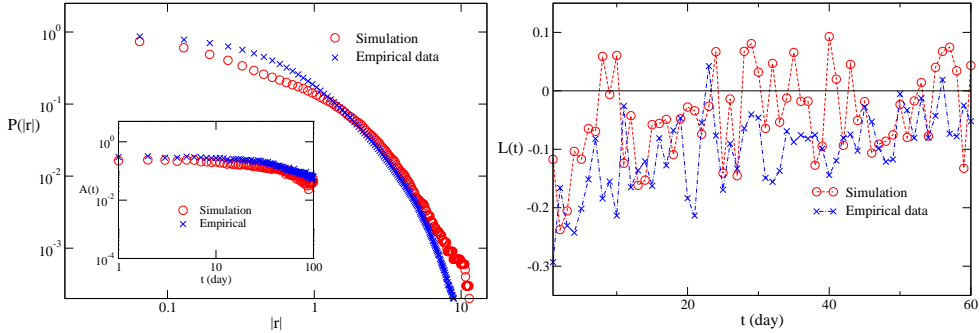

 \centering
\includegraphics[width=0.4\textwidth,clip]{new.eps}
\includegraphics[width=0.4\textwidth,clip]{Figure4a.eps}
\protect\caption{Comparison of the S\&P 500 components and the simulations: (a) The probability distribution functions of the absolute values of returns. The auto-correlation functions of volatilities are displayed in the inset. (b) The return-volatility correlation functions.
}
 \label{fig8} 
   \end{figure}

Our model reproduces the statistical features of the real stock markets. For instance, the simulation is compared with the daily price returns of the S\&P 500 components. To reduce the fluctuations, the calculations for the empirical data are averaged over all stocks. The probability distribution functions $P(|r(t)|)$ of the absolute values of returns are displayed in Fig.~\ref{fig8}(a), and the empirical fat tails are observed. The volatility clustering is characterized by the auto-correlation function of volatilities \cite{gop99}, which is defined in Eq.~(\ref{At}). As shown in the inset of figure Fig.~\ref{fig8}(a), $A(t)$ from the simulation is in agreement with that from the empirical data.

To describe how past returns affect future volatilities, we compute the return-volatility
correlation function $L(t)$ defined in Eq.~(\ref{L}).
As displayed in Fig.~\ref{fig8}(b), $L(t)$ from our simulation is consistent with that from empirical data.

\section{Summary}

\noindent
We first review several agent-based models and the new approaches to determine the key model parameters from historical market data.
Based on the agents' behaviors with heterogenous personal preferences and interactions, these models are successful
to explain the microscopic origination of the temporal and spatial correlations of the financial markets.
More specifically, the asymmetric trading and asymmetric herding are introduced to the agent-based modeling to understand the leverage and anti-leverage effects.
The asymmetric trading preference in volatile and stable markets is proposed to explain the non-local return-volatility correlation.
Finally, an agent-based model with the multi-level herding is constructed to simulate the sector structure.

We then present a novel paradigm combining the big-data analysis with the agent-based modeling. From internet query and stock market data, we extract the information driving forces, and develop an agent-based model to simulate the dynamic behaviors of the complex financial systems. The key parameters of the model are determined from the statistical properties of the information driving forces.
Our results provide a better understanding of the controlling effect of the information driving force on the complex financial system.
The ideology of the information driving force may be applied to the agent-based modeling of other open complex systems.

{\bf Acknowledgements:} This work was supported in part by NNSF of China under Grant Nos. 11375149 and 11505099.

\begin{small}

\end{small}

\end{document}